\documentclass[a4paper,noshowpacs,pra,preprint,preprintnumbers,superscriptaddress,longbibliography,floatfix]{revtex4-1}
\usepackage[utf8]{inputenc}
\usepackage{amsfonts,amsmath,amssymb}
\usepackage{graphicx}
\usepackage{bm} 
\usepackage{mathrsfs}
\usepackage{subfigure}
\usepackage{textcomp}
\usepackage{hyperref}
\usepackage[flushleft]{threeparttable}
\usepackage[per-mode=symbol]{siunitx}
\usepackage[version=3]{mhchem}
\graphicspath{{../figures/}}

\DeclareSIUnit\intensity{\watt\per\centi\meter\squared}
\DeclareSIUnit\fieldstrength{\volt\per\centi\meter}
\usepackage{xspace}

\DeclareUnicodeCharacter{2009}{\,}

\newcommand{\degree}{\ensuremath{^\circ}}%
\newcommand{\cost}{\ensuremath{\langle\cos^2\theta_\text{2D}\rangle}}
\newcommand{\costh}{\ensuremath{\langle\cos^2\theta \rangle}}

\newlength{\figwidth}
\newlength{\figwidthwide}
\let\orgautoref\autoref
\providecommand{\Autoref}{%
  \def\equationautorefname{Equation}%
  \def\figureautorefname{Figure}%
  \def\subfigureautorefname{Figure}%
  \orgautoref}
\renewcommand{\autoref}{%
  \def\equationautorefname{Eq.}%
  \def\figureautorefname{Fig.}%
  \def\subfigureautorefname{Fig.}%
  \orgautoref}

\usepackage{color}
\usepackage[normalem]{ulem}
\definecolor{darkgreen}{rgb}{0.0,0.7,0.0}

\newcommand{\csmon}{\ce{CS2}\xspace}
\newcommand{\csion}{\ce{CS2+}\xspace}

\begin{document}

\title{Alignment of the \ce{CS2} Dimer Embedded in Helium Droplets Induced by a Circularly Polarized Laser Pulse}
\author{James D.\ Pickering}
\author{Benjamin Shepperson}
\author{Lars Christiansen}
\author{Henrik Stapelfeldt}
\email{henriks@chem.au.dk}
\affiliation{Department of Chemistry, Aarhus University, Langelandsgade 140, 8000 Aarhus C}
\date{\today}
\begin{abstract}
	Dimers of carbon disulfide (\csmon) molecules embedded in helium nanodroplets are aligned using a moderately intense, \SI{160}{\pico\second}, non-resonant, circularly polarized laser pulse. It is shown that the intermolecular carbon-carbon (C-C) axis aligns along the axis perpendicular to the polarization plane of the alignment laser pulse. The degree of alignment, quantified by \cost, is determined from the emission directions of recoiling \csion fragment ions, created when an intense \SI{40}{\femto\second} probe laser pulse doubly ionizes the dimers. Here, \ensuremath{\theta_\text{2D}} is the projection of the angle between the C-C axis on the 2D ion detector and the normal to the polarization plane. \cost~is measured as a function of the alignment laser intensity and the results agree well with \cost~calculated for gas-phase \csmon dimers with a rotational temperature of \SI{0.4}{\kelvin}.
\end{abstract}
\maketitle
\section{Introduction}
Recently, it has been shown that techniques used to align gas-phase samples of molecules based on the application of moderately intense laser pulses can be transferred to molecules inside helium nanodroplets~\cite{pentlehner_impulsive_2013,pentlehner_laser-induced_2013,shepperson_laser-induced_2017,shepperson_strongly_2017,chatterley_three-dimensional_2017}. This creates several new opportunities. Firstly, time-resolved imaging of molecular rotation, induced and probed by fs laser pulses makes it possible to investigate how rotational quantum coherence, angular momentum, and energy are influenced by a dissipative environment~\cite{pentlehner_impulsive_2013,shepperson_laser-induced_2017,shepperson_observation_2018,ramakrishna_intense_2005,vieillard_field-free_2008,lindgren_excitation_2013,stickler_rotational_2018,schmidt_rotation_2015,lemeshko_quasiparticle_2017}. Secondly, the \SI{0.4}{\kelvin} temperature of He droplets is shared with the molecules residing inside the droplets~\cite{toennies_superfluid_2004}. Such a low temperature is highly beneficial for creating very strong alignment -- not just for small molecules~\cite{shepperson_strongly_2017} but also for larger species~\cite{chatterley_long_2018}, which can otherwise be difficult to effectively cool in the gas phase~\cite{meijer_laser_1990}. Thirdly, the cold and dissipative environment of the He droplet provides unique opportunities for creating a variety of complexes of molecules and atoms~\cite{choi_infrared_2006,yang_helium_2012}. Recently, it was shown that laser-induced alignment can also be applied efficiently to molecular complexes inside He droplets~\cite{pickering_femtosecond_2018,pickering_alignment_2018}. The high degrees of alignment made it possible to determine the structure of the molecular complexes by means of Coulomb explosion imaging induced by a high-intensity femtosecond laser pulse. Here we present a follow-up to our previous study of the \csmon dimer~\cite{pickering_alignment_2018}, examining the alignment behaviour of the dimer in more detail.

Alignment of molecules using linearly polarized laser pulses results in the confinement of the most polarizable axis of a molecule to the polarization axis of the aligning laser field and creates one-dimensional alignment~\cite{stapelfeldt_colloquium:_2003}. This has been the most common use of laser-induced alignment in many prior studies. The most polarizable axis provides a `handle', which can be used to hold the molecule in advantageous spatial configurations using the alignment laser field~\cite{burt_communication_2018}. However, it is also possible to align a molecule using a circularly (or elliptically) polarized laser field, which provides different handles and allows the molecule to be held in other spatial configurations than are possible with a linearly polarized field. Specifically, using a circularly polarized alignment field results in the alignment of the least polarizable molecular axis to the propagation axis of the alignment laser field~\cite{smeenk_molecular_2013}. This can be equivalently thought of as alignment of the most polarizable molecular plane to the polarization plane.

Our previous study of the \csmon dimer solvated in He droplets~\cite{pickering_alignment_2018} showed that it was possible to form, align, and image \csmon dimers inside He droplets. Analysis of the recoil directions of \csion ions produced by Coulomb explosion of aligned \csmon dimers showed that the structure of the \csmon dimer in the He droplets was cross-shaped with $D_{2d}$ symmetry - which is also the structure in the gas phase~\cite{moazzen-ahmadi_spectroscopy_2013,rezaei_spectroscopic_2011}. Central to that study was the alignment of the \csmon dimer (a symmetric top rotor) using a circularly polarized laser pulse. In the current work, we perform a detailed study of the intensity-dependent alignment of the \csmon dimer in He droplets using a circularly polarized laser pulse. The alignment pulse duration of 160 ps places the alignment dynamics in the (quasi-) adiabatic regime. To our knowledge very few studies so far have addressed alignment of symmetric top molecules with a circularly polarized alignment pulse, and only in the nonadiabatic limit using femtosecond alignment pulses~\cite{smeenk_molecular_2013,smeenk_alignment_2014}.

\section{Background}
\subsection{Experimental Methods}
The experimental setup used to align and image the \csmon dimers has already been described in detail elsewhere~\cite{shepperson_strongly_2017,pickering_alignment_2018}, with only relevant details reiterated here. The \csmon molecules were embedded in He droplets consisting of around 8000 He atoms on average. The droplets are passed through a pickup cell containing \csmon vapor, and the partial pressure of \csmon was adjusted to a regime where substantial amounts of dimers (but minimal amounts of larger clusters) were formed. This is defined as the dimer-doping-condition - in line with previous work~\cite{pickering_alignment_2018}. The doped droplet beam is then intersected at \SI{90}{\degree} by two focussed laser beams from the same Ti-Sapphire laser system. One beam consists of \SI{160}{\pico\second} (FWHM) pulses ($\lambda_{\text{center}}$ = \SI{800}{\nano\metre}) and is used to align the \csmon dimers. The other beam consists of \SI{40}{\femto\second} (FWHM) pulses ($\lambda_{\text{center}}$ = \SI{800}{\nano\metre}), and is used to Coulomb explode the dimers. The intensity of the probe pulses was \SI{3d14}{\intensity}, and the intensity of the alignment pulses was varied, up to a maximum value of \SI{8d11}{\intensity}. The temporal overlap of the two pulses was such that the probe pulse arrived at the peak of the alignment pulse. The nascent \csion ions were projected onto a 2D imaging detector using a VMI spectrometer, and ion images were recorded.

\begin{figure*}
	\centering
	\includegraphics{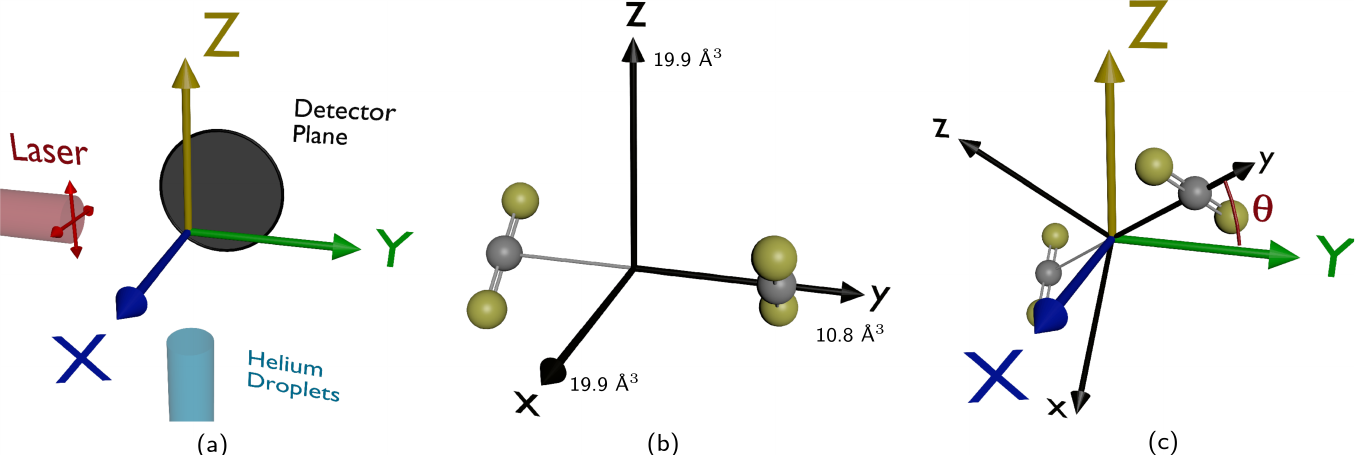}
	\caption{(a) Sketch of the space-fixed (XYZ) coordinate system showing the propagation direction of the laser beams, the He droplet beam, and the detector. (b) Sketch of the molecular (xyz) coordinate system for the \csmon dimer in its ground state geometry. Polarizability elements of the dimer (in units of \si{\angstrom\cubed}) are annotated onto each molecular axis (see text). (c) Sketch of the combined space-fixed (color) and molecular (black) coordinate systems, with the angle $\theta$ (between the y and the Y axes) shown in red.}\label{fig:coordinate_systems}
\end{figure*}

Before discussing the details of laser-induced alignment, it is important to define the molecular (x, y, z) and space-fixed (X, Y, Z) coordinate systems. \Autoref{fig:coordinate_systems}\,(a) shows the space-fixed coordinate system. The laser beams propagate along the Y axis, such that the XZ plane is the polarization plane, and the He droplet beam propagates along the Z axis. Ions are accelerated along the X axis (the time-of-flight axis) towards a detector placed in the YZ plane. \Autoref{fig:coordinate_systems}\,(b) shows the structure of the \csmon dimer in its ground state geometry. The \csmon dimer is a prolate symmetric top with a 90{\degree} angle between the two monomers, and in this case the symmetric top axis (the C-C axis) is placed along the y axis. The polarizability elements of the dimer are annotated on each molecular axis. In this case, the polarizability elements of the dimer were calculated simply by addition of the polarizability elements of the two monomers. This method neglects any orbital interaction between the two monomers, but will not qualitatively change the polarizability elements of the dimer, so is a reasonable first approximation to the true polarizability.

\subsection{Alignment Theory}
The methodology and theory for alignment of molecules induced by strong laser fields has been described in several reviews~\cite{stapelfeldt_colloquium:_2003,seideman_nonadiabatic_2005,fleischer_molecular_2012}. Here we specifically consider the alignment of a symmetric top molecule using a circularly polarized laser pulse~\cite{smeenk_molecular_2013,smeenk_alignment_2014}. The interaction potential $\hat{V}(t)$ of a symmetric top molecule with a circularly polarized alignment field is given by~\cite{larsen_laser_2000}:
\begin{equation}\label{eq:alignment_potential}
	\hat{V}(t) = -\frac{1}{4} E_0^2(t)[(\alpha_\perp - \alpha_\parallel)\cos^2(\theta) + \alpha_\parallel + \alpha_\perp]
\end{equation}
where $E_0(t)$ is the electric field envelope of the alignment laser field; $\alpha_\parallel$ is the polarizability of the molecule parallel to the symmetric top axis; $\alpha_\perp$ is the polarizability perpendicular to the symmetric top axis; and $\theta$ is the angle between the symmetric top axis and the normal to the polarization plane of the laser field. The alignment behaviour is determined by the sign of the polarizability anisotropy, $\alpha_\parallel$ - $\alpha_\perp$. If $\alpha_\parallel < \alpha_\perp$, which is typically the case for oblate symmetric top molecules like benzene (\ce{C_6H_6}), ammonia (\ce{NH_3}) or boron trifluoride (\ce{BF_3}), the potential is at a minimum when $\theta = \SI{0}{\degree}$ and the interaction between the laser field and the molecule should lead to alignment of the symmetric top axis along the normal vector of the polarization plane, i.e. the propagation direction of the laser field. This has been illustrated for benzene molecules using nonadiabatic alignment induced by a 250 fs alignment pulse~\cite{smeenk_molecular_2013}. In contrast, if $\alpha_\perp < \alpha_\parallel$, which is typically the case for prolate symmetric top molecules like methyl iodide (\ce{CH3I}) or acetonitrile (\ce{CH_3CN}), or for linear molecules, then the potential is at a minimum when $\theta = \SI{90}{\degree}$. In this case the symmetric top axis will be confined to (but randomly oriented in) the polarization plane.

For the \csmon dimer $\alpha_\parallel < \alpha_\perp$ (\autoref{fig:coordinate_systems}\,(b)), although it is a prolate symmetric top. Thus, the expectation is that a circularly polarized alignment field will align the C-C axis (the y axis) along the propagation direction of the laser beam (the Y axis). A sketch of the combined space-fixed and molecule-fixed coordinate systems, and an illustration of the angle $\theta$, is given in~\autoref{fig:coordinate_systems}\,(c).

\subsection{Computational Methods}
Recently, angulon theory~\cite{lemeshko_quasiparticle_2017,schmidt_rotation_2015} was used to rationalize that for the \SI{160}{\pico\second} alignment pulse used here, alignment of \ce{I2}, 1,4-diiodobenzene, and 1,4-dibromobenzene molecules inside He droplets is well-described as alignment of gas-phase molecules with an effective rotational constant~\cite{shepperson_strongly_2017}. Qualitatively, this can be understood using the fact that the turn-on time of the alignment pulse is longer than the timescale of the He-He interactions and the He-molecule interactions. Therefore, a molecule and its solvation shell of He atoms has time to adjust as the laser field increases in strength. The \ce{CS2} dimer is expected to behave similarly and thus we apply gas-phase alignment theory to calculate the degree of alignment of the dimer induced by the circularly polarized alignment pulse. In practice, the calculations were performed using an alignment calculator previously developed in our group~\cite{sondergaard_nonadiabatic_2017}. This calculator solves the time-dependent rotational Schr{\"o}dinger equation and returns the degree of alignment expressed as \costh or \cost. Here, $\theta_{\text{2D}}$ is the angle between the projection of the molecular C-C axis on the detector plane and the laser propagation axis. Calculation of \cost allows direct comparison of the calculations with the experimental data.

Strictly, the alignment calculator can only simulate alignment of linear or symmetric top molecules with $\alpha_\parallel > \alpha_\perp$ using linearly polarized alignment fields where the interaction potential is given by~\cite{friedrich_alignment_1995}:

\begin{equation}\label{eq:alignment_potential_linear_molecule}
	\hat{V}(t) = -\frac{1}{4} E_0^2(t)[(\alpha_\parallel - \alpha_\perp)\cos^2(\theta) + \alpha_\perp]
\end{equation}

Comparison of \autoref{eq:alignment_potential} and \autoref{eq:alignment_potential_linear_molecule} shows that for a symmetric top molecule where $\alpha_\perp > \alpha_\parallel$ the angle-dependent part of the interaction potential induced by a circularly polarized field is the same as that for a linear molecule induced by a linearly polarized field. For the \csmon dimer $\alpha_\perp > \alpha_\parallel$. This means that the alignment calculator can be used to calculate the degree of alignment of the \csmon dimer induced by the circularly polarized field. We note that experimentally the peak intensity, $I_0$, is determined. For a circularly polarized field the relation to the amplitude of the electric field, $E_0$, is given by $I_0 = \varepsilon_0 c E_0^2$.

A detailed description of the alignment calculator can be found in reference~\cite{sondergaard_understanding_2016}, with only relevant parameters mentioned here. Calculations were performed at a variety of ensemble temperatures (see later), using the experimentally measured Gaussian focal spot-sizes for both the alignment ($\omega_0 =$ \SI{35}{\micro\metre}) and probe ($\omega_0 =$ \SI{25}{\micro\metre}) beams. The degree of alignment was calculated using both the polarizability anisotropy determined by addition of the monomer polarizabilities ($\alpha_\parallel =$\SI{10.8}{\angstrom\cubed}, $\alpha_\perp =$\SI{19.9}{\angstrom\cubed}), and that calculated using density-functional theory (DFT - wB97XD/aug-pcseg-2) ($\alpha_\parallel =$\SI{12.0}{\angstrom\cubed}, $\alpha_\perp =$\SI{17.5}{\angstrom\cubed})~\cite{pickering_alignment_2018}. The effective rotational constant of a single \csmon molecule inside He droplets has been calculated to be $0.3B_0$~\cite{zillich_private_2018}, where $B_0$ is the gas-phase rotational constant (\SI{0.109}{\centi\meter^{-1}}). A similar reduction is assumed for the \csmon dimer, thus the rotational constants used in the calculation are $A =$ \SI{0.0182}{\centi\meter^{-1}}, $B = C =$ \SI{0.01281}{\centi\meter^{-1}}, where $A$ is calculated around the molecular y axis, and $B$ and $C$ are calculated around the molecular x and z axes respectively. We note that in the (quasi-) adiatbatic limit defined by the \SI{160}{\pico\second} pulses the degree of alignment is very insensitive to the rotational constant, being essentially only determined by the polarizability anisotropy, rotational temperature of the molecular ensemble and the intensity of the alignment pulse~\cite{seideman_dynamics_2001}.

\section{Results}

\begin{figure*}
	\centering
	\includegraphics[width=\figwidthwide]{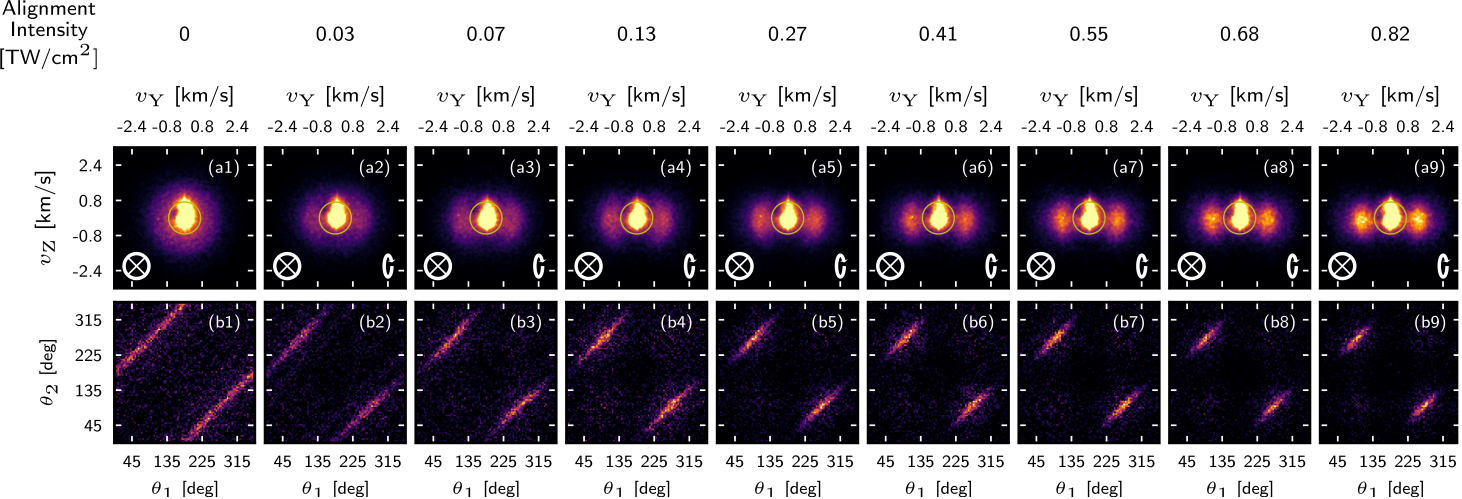}
	\caption{(a1)-(a9) \csion ion images and (b1)-(b9) corresponding angular covariance maps calculated using ions outside the annotated yellow circles in row (a). The polarization state of the probe [alignment] laser is shown in the lower left [right] corner of each ion image. All data was recorded under the dimer-doping-condition. The data in each column (1)-(9) was recorded at a different alignment intensity, annotated above each column. }\label{fig:image_series}
\end{figure*}

Each column of~\autoref{fig:image_series} shows both a \csion ion image (row (a)), and corresponding angular covariance map (row (b)). The angular covariance map~\cite{hansen_control_2012} was calculated over all ion hits outside of the annotated yellow circle on the ion image - this ensures that the covariance map is calculated using ion events arising from the Coulomb explosion of \csmon dimers - as discussed in previous work~\cite{pickering_alignment_2018}. The Coulomb explosion channel considered here is the one resulting from repulsion of two \csion ions created when each of the two \csmon molecules in the dimer are singly ionized by the probe laser pulse. The ion signal inside the yellow circle arises from single ionization of \csmon monomers. In all images, the probe pulse was linearly polarized along the X axis (shown in the lower left corner of each ion image), and the alignment pulse was circularly polarized in the XZ plane (shown in the lower right corner of each ion image). The alignment intensity used to record each image is shown above each column, and was incremented from \SI{0}{\tera\intensity} (column 1) to \SI{0.82}{\tera\intensity} (column 9).

The ion signal outside the annotated yellow circle in images (a1)-(a9) of~\autoref{fig:image_series} moves from being isotropic (a1), to being strongly confined along the Y axis (a9) as the alignment intensity is increased from zero to its maximum value. This confinement is also reflected in the corresponding angular covariance maps (b1)-(b9), where the covariance signal changes from being a uniform line centered at $\theta_2 = \theta_1 +$\SI{180}{\degree} or at $\theta_2 = \theta_1 -$\SI{180}{\degree} (panel (b1)), to a confined island centered at (\SI{90}{\degree},\SI{270}{\degree}), and the equivalent island at (\SI{270}{\degree},\SI{90}{\degree}) (panel (b9)). It is clear from both the ion images and covariance maps that the degree of alignment increases as the alignment intensity is increased from column (1) to column (9), and that the degree of alignment appears to saturate at around column (5) - increasing the intensity beyond this level does not seem to dramatically increase the observed degree of alignment. This can be quantified by calculating \cost \, - but it is useful to first consider the explosion process in He droplets in more detail.

Prior studies of aligned molecules in He droplets have shown that the recoil trajectories of the nascent ion fragments following Coulomb explosion are distorted due to scattering off the He atoms as they exit the droplet. This results in Non-Axial Recoil (NAR), where the recoil trajectories of the nascent ions no longer directly reflect the alignment of the molecular axis they recoiled from. In studies of laser-induced alignment inside He droplets, calculating \cost\, from an ion image without correcting for NAR leads to an underestimate of the true degree of alignment. However, the effect of the NAR can be determined from the correlation between the fragment ions (here the \csion ions) using covariance analysis, and deconvoluted from the fragment angular distributions, allowing the true degree of alignment to be obtained~\cite{christensen_deconvoluting_2016}. This has been implemented in the study of aligned molecules in He droplets previously~\cite{shepperson_strongly_2017}.

\begin{figure}
	\centering
	\includegraphics[width=\figwidth]{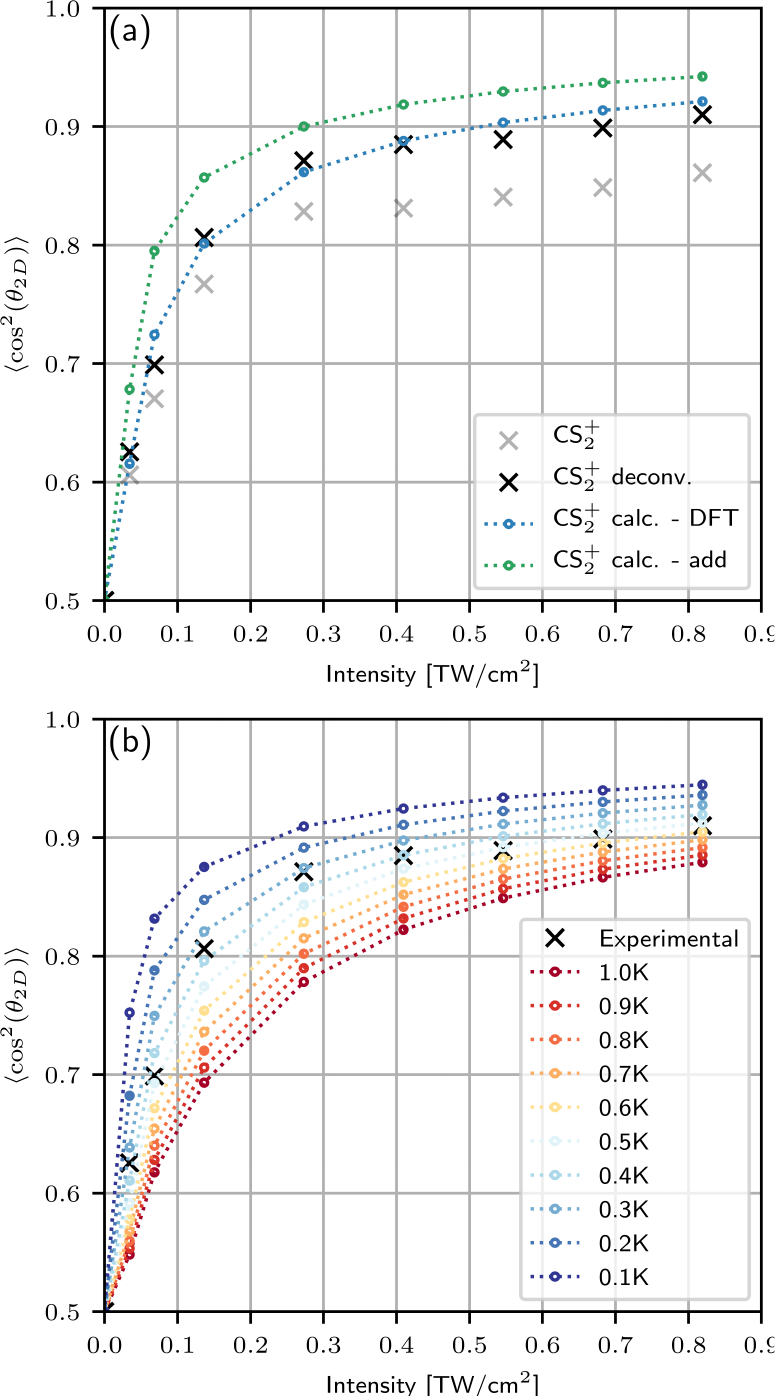}
	\caption{Experimental (crosses) and simulated (circles and dashed line) degree of alignment of the \csmon dimer as a function of alignment intensity. The alignment was induced adiabatically using a circularly polarized laser pulse. Panel (a) shows a comparison of the experimental degree of alignment with the simulated degree of alignment calculated at \SI{0.4}{\kelvin}, using two different polarizability anisotropies (see text). Panel (b) shows a comparison of the experimental degree of alignment with the simulated degree of alignment calculated at a variety of different ensemble temperatures (see text).}\label{fig:alignment_traces}
\end{figure}

\Autoref{fig:alignment_traces}\,(a) shows the experimentally measured (crosses) and the simulated (circular markers and dashed line) degree of alignment (\cost) of the \csmon dimer as a function of the alignment pulse intensity. The grey crosses represent the degree of alignment determined from the \csion ion images without any correction for NAR, whereas the black crosses show the same measurements, corrected for NAR. The blue simulated trace was calculated using a polarizability anisotropy calculated using DFT assuming an inter-monomer separation of \SI{3.5}{\angstrom} (such that $\Delta \alpha =$\SI{5.5}{\angstrom\cubed}), whereas the green trace was calculated using a polarizability anistropy calculated by addition of the polarizability tensors of the two \csmon monomers (such that $\Delta \alpha =$\SI{9.1}{\angstrom\cubed}). \Autoref{fig:alignment_traces}\,(b) again shows the experimentally measured degree of alignment, corrected for NAR (crosses), together with the simulated degree of alignment calculated at a variety of different ensemble temperatures. For all the different ensemble temperatures, the polarizability anisotropy was chosen to be that which was calculated using DFT.

\section{Discussion}
Examining~\autoref{fig:alignment_traces}\,(a), several observations can be made. Firstly, it is clear that the observed degree of alignment rapidly increases as the alignment intensity is raised, before levelling out at higher intensities. Secondly, the deconvoluted data (black crosses) shows a higher degree of alignment than the data obtained directly from the ion images (grey crosses), as is expected due to the NAR in He droplets. Accordingly, the deconvoluted data is a better match to the simulated traces than the non-deconvoluted data. Additionally, the simulated trace that best matches the experimental data is the one employing a polarisability anistropy calculated by DFT methods, rather than that calculated via addition of the polarizability tensors of the two monomers.

The polarizability anisotropy calculated by DFT compared to the anisotropy calculated via addition of the polarizability tensors of the two monomers. This implies that there is some electronic (orbital) interaction between the two \csmon units in the dimer, as the simple addition of polarizability elements neglects this. The electronic interaction lowers the overall polarizability anisotropy as the electron density in the region between the two \csmon molecules is increased. This suggests that the interaction between the two \csmon monomers has some orbital character - in addition to the expected (non-orbital) van der Waals interaction~\cite{hobza_noncovalent_2010}. The sensitivity to the polarizability anisotropy in this way illustrates the potential utility of laser-induced alignment as a diagnostic tool for studying non-covalent interactions and that it may be possible to gain some deeper insight into the electronic structure of such complexes with further study.

Turning to \autoref{fig:alignment_traces}\,(b), it appears that the experimental data corresponds well to the simulated data using an ensemble temperature of between 0.3 and \SI{0.5}{\kelvin}. This agreement implies that the alignment of the \csmon dimers in the He droplets is well described by alignment of a \SI{0.4}{\kelvin} ensemble of gas-phase \csmon dimers. This is completely in line with the expected temperature of the dimers inside the He droplets~\cite{toennies_superfluid_2004}.

\section{Conclusion}

We have shown that the C-C axis of \csmon dimers embedded in He droplets can be tightly confined along the propagation direction of a circularly polarized \SI{160}{\pico\second} laser pulse. The degree of this one-dimensional alignment was measured as a function of the intensity of the alignment pulse by Coulomb explosion imaging with an intense \SI{40}{\femto\second} probe pulse. It was found that the measured degree of alignment, \cost, as a function of intensity), matched well with \cost calculated for a sample of \csmon dimers with an adjusted rotational constant and a temperature of \SI{0.4}{\kelvin}. As such, our results corroborate and generalize recent findings on \ce{I_2} molecules in He droplets where the experimentally recorded degree of one-dimensional alignment, induced by a linearly polarized laser field, also agreed well with calculations of \cost\, for a \SI{0.4}{\kelvin} ensemble of \ce{I_2} molecules with an adjusted rotational constant. Since molecules and complexes of molecules embedded in He droplets should, in general, thermalize to the \SI{0.4}{\kelvin} environment we believe that the high degree of alignment enabled by the low temperature can be obtained for a wide variety of species. This will prove advantageous for applications that benefit from such rigorous control of the molecular frame, including diffractive imaging with (soft) x-rays~\cite{kupper_x-ray_2014,gomez_shapes_2014,tanyag_communication:_2015,rupp_coherent_2017}, Coulomb explosion imaging~\cite{pickering_alignment_2018}, and electron diffraction~\cite{he_self_2017,zhang_electron_2014}.


\begin{thebibliography}{42}%
\makeatletter
\providecommand \@ifxundefined [1]{%
 \@ifx{#1\undefined}
}%
\providecommand \@ifnum [1]{%
 \ifnum #1\expandafter \@firstoftwo
 \else \expandafter \@secondoftwo
 \fi
}%
\providecommand \@ifx [1]{%
 \ifx #1\expandafter \@firstoftwo
 \else \expandafter \@secondoftwo
 \fi
}%
\providecommand \natexlab [1]{#1}%
\providecommand \enquote  [1]{``#1''}%
\providecommand \bibnamefont  [1]{#1}%
\providecommand \bibfnamefont [1]{#1}%
\providecommand \citenamefont [1]{#1}%
\providecommand \href@noop [0]{\@secondoftwo}%
\providecommand \href [0]{\begingroup \@sanitize@url \@href}%
\providecommand \@href[1]{\@@startlink{#1}\@@href}%
\providecommand \@@href[1]{\endgroup#1\@@endlink}%
\providecommand \@sanitize@url [0]{\catcode `\\12\catcode `\$12\catcode
  `\&12\catcode `\#12\catcode `\^12\catcode `\_12\catcode `\%12\relax}%
\providecommand \@@startlink[1]{}%
\providecommand \@@endlink[0]{}%
\providecommand \url  [0]{\begingroup\@sanitize@url \@url }%
\providecommand \@url [1]{\endgroup\@href {#1}{\urlprefix }}%
\providecommand \urlprefix  [0]{URL }%
\providecommand \Eprint [0]{\href }%
\providecommand \doibase [0]{http://dx.doi.org/}%
\providecommand \selectlanguage [0]{\@gobble}%
\providecommand \bibinfo  [0]{\@secondoftwo}%
\providecommand \bibfield  [0]{\@secondoftwo}%
\providecommand \translation [1]{[#1]}%
\providecommand \BibitemOpen [0]{}%
\providecommand \bibitemStop [0]{}%
\providecommand \bibitemNoStop [0]{.\EOS\space}%
\providecommand \EOS [0]{\spacefactor3000\relax}%
\providecommand \BibitemShut  [1]{\csname bibitem#1\endcsname}%
\let\auto@bib@innerbib\@empty
\bibitem [{\citenamefont {Pentlehner}\ \emph
  {et~al.}(2013{\natexlab{a}})\citenamefont {Pentlehner}, \citenamefont
  {Nielsen}, \citenamefont {Slenczka}, \citenamefont {Mølmer},\ and\
  \citenamefont {Stapelfeldt}}]{pentlehner_impulsive_2013}%
  \BibitemOpen
  \bibfield  {author} {\bibinfo {author} {\bibfnamefont {D.}~\bibnamefont
  {Pentlehner}}, \bibinfo {author} {\bibfnamefont {J.~H.}\ \bibnamefont
  {Nielsen}}, \bibinfo {author} {\bibfnamefont {A.}~\bibnamefont {Slenczka}},
  \bibinfo {author} {\bibfnamefont {K.}~\bibnamefont {Mølmer}}, \ and\
  \bibinfo {author} {\bibfnamefont {H.}~\bibnamefont {Stapelfeldt}},\
  }\bibfield  {title} {\enquote {\bibinfo {title} {Impulsive {Laser} {Induced}
  {Alignment} of {Molecules} {Dissolved} in {Helium} {Nanodroplets}},}\ }\href
  {\doibase 10.1103/PhysRevLett.110.093002} {\bibfield  {journal} {\bibinfo
  {journal} {Phys. Rev. Lett.}\ }\textbf {\bibinfo {volume} {110}},\ \bibinfo
  {pages} {093002} (\bibinfo {year} {2013}{\natexlab{a}})}\BibitemShut
  {NoStop}%
\bibitem [{\citenamefont {Pentlehner}\ \emph
  {et~al.}(2013{\natexlab{b}})\citenamefont {Pentlehner}, \citenamefont
  {Nielsen}, \citenamefont {Christiansen}, \citenamefont {Slenczka},\ and\
  \citenamefont {Stapelfeldt}}]{pentlehner_laser-induced_2013}%
  \BibitemOpen
  \bibfield  {author} {\bibinfo {author} {\bibfnamefont {D.}~\bibnamefont
  {Pentlehner}}, \bibinfo {author} {\bibfnamefont {J.~H.}\ \bibnamefont
  {Nielsen}}, \bibinfo {author} {\bibfnamefont {L.}~\bibnamefont
  {Christiansen}}, \bibinfo {author} {\bibfnamefont {A.}~\bibnamefont
  {Slenczka}}, \ and\ \bibinfo {author} {\bibfnamefont {H.}~\bibnamefont
  {Stapelfeldt}},\ }\bibfield  {title} {\enquote {\bibinfo {title}
  {Laser-induced adiabatic alignment of molecules dissolved in helium
  nanodroplets},}\ }\href {\doibase 10.1103/PhysRevA.87.063401} {\bibfield
  {journal} {\bibinfo  {journal} {Phys. Rev. A}\ }\textbf {\bibinfo {volume}
  {87}},\ \bibinfo {pages} {063401} (\bibinfo {year}
  {2013}{\natexlab{b}})}\BibitemShut {NoStop}%
\bibitem [{\citenamefont {Shepperson}\ \emph
  {et~al.}(2017{\natexlab{a}})\citenamefont {Shepperson}, \citenamefont
  {Søndergaard}, \citenamefont {Christiansen}, \citenamefont {Kaczmarczyk},
  \citenamefont {Zillich}, \citenamefont {Lemeshko},\ and\ \citenamefont
  {Stapelfeldt}}]{shepperson_laser-induced_2017}%
  \BibitemOpen
  \bibfield  {author} {\bibinfo {author} {\bibfnamefont {B.}~\bibnamefont
  {Shepperson}}, \bibinfo {author} {\bibfnamefont {A.~A.}\ \bibnamefont
  {Søndergaard}}, \bibinfo {author} {\bibfnamefont {L.}~\bibnamefont
  {Christiansen}}, \bibinfo {author} {\bibfnamefont {J.}~\bibnamefont
  {Kaczmarczyk}}, \bibinfo {author} {\bibfnamefont {R.~E.}\ \bibnamefont
  {Zillich}}, \bibinfo {author} {\bibfnamefont {M.}~\bibnamefont {Lemeshko}}, \
  and\ \bibinfo {author} {\bibfnamefont {H.}~\bibnamefont {Stapelfeldt}},\
  }\bibfield  {title} {\enquote {\bibinfo {title} {Laser-{Induced} {Rotation}
  of {Iodine} {Molecules} in {Helium} {Nanodroplets}: {Revivals} and {Breaking}
  {Free}},}\ }\href {\doibase 10.1103/PhysRevLett.118.203203} {\bibfield
  {journal} {\bibinfo  {journal} {Phys. Rev. Lett.}\ }\textbf {\bibinfo
  {volume} {118}},\ \bibinfo {pages} {203203} (\bibinfo {year}
  {2017}{\natexlab{a}})}\BibitemShut {NoStop}%
\bibitem [{\citenamefont {Shepperson}\ \emph
  {et~al.}(2017{\natexlab{b}})\citenamefont {Shepperson}, \citenamefont
  {Chatterley}, \citenamefont {Søndergaard}, \citenamefont {Christiansen},
  \citenamefont {Lemeshko},\ and\ \citenamefont
  {Stapelfeldt}}]{shepperson_strongly_2017}%
  \BibitemOpen
  \bibfield  {author} {\bibinfo {author} {\bibfnamefont {B.}~\bibnamefont
  {Shepperson}}, \bibinfo {author} {\bibfnamefont {A.~S.}\ \bibnamefont
  {Chatterley}}, \bibinfo {author} {\bibfnamefont {A.~A.}\ \bibnamefont
  {Søndergaard}}, \bibinfo {author} {\bibfnamefont {L.}~\bibnamefont
  {Christiansen}}, \bibinfo {author} {\bibfnamefont {M.}~\bibnamefont
  {Lemeshko}}, \ and\ \bibinfo {author} {\bibfnamefont {H.}~\bibnamefont
  {Stapelfeldt}},\ }\bibfield  {title} {\enquote {\bibinfo {title} {Strongly
  aligned molecules inside helium droplets in the near-adiabatic regime},}\
  }\href {\doibase 10.1063/1.4983703} {\bibfield  {journal} {\bibinfo
  {journal} {J. Chem. Phys.}\ }\textbf {\bibinfo {volume} {147}},\ \bibinfo
  {pages} {013946} (\bibinfo {year} {2017}{\natexlab{b}})}\BibitemShut
  {NoStop}%
\bibitem [{\citenamefont {Chatterley}\ \emph {et~al.}(2017)\citenamefont
  {Chatterley}, \citenamefont {Shepperson},\ and\ \citenamefont
  {Stapelfeldt}}]{chatterley_three-dimensional_2017}%
  \BibitemOpen
  \bibfield  {author} {\bibinfo {author} {\bibfnamefont {A.~S.}\ \bibnamefont
  {Chatterley}}, \bibinfo {author} {\bibfnamefont {B.}~\bibnamefont
  {Shepperson}}, \ and\ \bibinfo {author} {\bibfnamefont {H.}~\bibnamefont
  {Stapelfeldt}},\ }\bibfield  {title} {\enquote {\bibinfo {title}
  {Three-{Dimensional} {Molecular} {Alignment} {Inside} {Helium}
  {Nanodroplets}},}\ }\href {\doibase 10.1103/PhysRevLett.119.073202}
  {\bibfield  {journal} {\bibinfo  {journal} {Phys. Rev. Lett.}\ }\textbf
  {\bibinfo {volume} {119}},\ \bibinfo {pages} {073202} (\bibinfo {year}
  {2017})}\BibitemShut {NoStop}%
\bibitem [{\citenamefont {Shepperson}\ \emph {et~al.}(2018)\citenamefont
  {Shepperson}, \citenamefont {Chatterley}, \citenamefont {Christiansen},
  \citenamefont {Søndergaard},\ and\ \citenamefont
  {Stapelfeldt}}]{shepperson_observation_2018}%
  \BibitemOpen
  \bibfield  {author} {\bibinfo {author} {\bibfnamefont {B.}~\bibnamefont
  {Shepperson}}, \bibinfo {author} {\bibfnamefont {A.~S.}\ \bibnamefont
  {Chatterley}}, \bibinfo {author} {\bibfnamefont {L.}~\bibnamefont
  {Christiansen}}, \bibinfo {author} {\bibfnamefont {A.~A.}\ \bibnamefont
  {Søndergaard}}, \ and\ \bibinfo {author} {\bibfnamefont {H.}~\bibnamefont
  {Stapelfeldt}},\ }\bibfield  {title} {\enquote {\bibinfo {title} {Observation
  of rotational revivals for iodine molecules in helium droplets using a
  near-adiabatic laser pulse},}\ }\href {\doibase 10.1103/PhysRevA.97.013427}
  {\bibfield  {journal} {\bibinfo  {journal} {Phys. Rev. A}\ }\textbf {\bibinfo
  {volume} {97}},\ \bibinfo {pages} {013427} (\bibinfo {year}
  {2018})}\BibitemShut {NoStop}%
\bibitem [{\citenamefont {Ramakrishna}\ and\ \citenamefont
  {Seideman}(2005)}]{ramakrishna_intense_2005}%
  \BibitemOpen
  \bibfield  {author} {\bibinfo {author} {\bibfnamefont {S.}~\bibnamefont
  {Ramakrishna}}\ and\ \bibinfo {author} {\bibfnamefont {T.}~\bibnamefont
  {Seideman}},\ }\bibfield  {title} {\enquote {\bibinfo {title} {Intense
  {Laser} {Alignment} in {Dissipative} {Media} as a {Route} to {Solvent}
  {Dynamics}},}\ }\href {\doibase 10.1103/PhysRevLett.95.113001} {\bibfield
  {journal} {\bibinfo  {journal} {Phys. Rev. Lett.}\ }\textbf {\bibinfo
  {volume} {95}},\ \bibinfo {pages} {113001} (\bibinfo {year}
  {2005})}\BibitemShut {NoStop}%
\bibitem [{\citenamefont {Vieillard}\ \emph {et~al.}(2008)\citenamefont
  {Vieillard}, \citenamefont {Chaussard}, \citenamefont {Sugny}, \citenamefont
  {Lavorel},\ and\ \citenamefont {Faucher}}]{vieillard_field-free_2008}%
  \BibitemOpen
  \bibfield  {author} {\bibinfo {author} {\bibfnamefont {T.}~\bibnamefont
  {Vieillard}}, \bibinfo {author} {\bibfnamefont {F.}~\bibnamefont
  {Chaussard}}, \bibinfo {author} {\bibfnamefont {D.}~\bibnamefont {Sugny}},
  \bibinfo {author} {\bibfnamefont {B.}~\bibnamefont {Lavorel}}, \ and\
  \bibinfo {author} {\bibfnamefont {O.}~\bibnamefont {Faucher}},\ }\bibfield
  {title} {\enquote {\bibinfo {title} {Field-free molecular alignment of
  {CO}$_2$ mixtures in presence of collisional relaxation},}\ }\href {\doibase
  10.1002/jrs.1976} {\bibfield  {journal} {\bibinfo  {journal} {J. Raman
  Spectrosc.}\ }\textbf {\bibinfo {volume} {39}},\ \bibinfo {pages} {694--699}
  (\bibinfo {year} {2008})}\BibitemShut {NoStop}%
\bibitem [{\citenamefont {Lindgren}\ and\ \citenamefont
  {Kiljunen}(2013)}]{lindgren_excitation_2013}%
  \BibitemOpen
  \bibfield  {author} {\bibinfo {author} {\bibfnamefont {J.}~\bibnamefont
  {Lindgren}}\ and\ \bibinfo {author} {\bibfnamefont {T.}~\bibnamefont
  {Kiljunen}},\ }\bibfield  {title} {\enquote {\bibinfo {title} {Excitation of
  rotons in parahydrogen crystals: {The} laser-induced-molecular-alignment
  mechanism},}\ }\href {\doibase 10.1103/PhysRevA.88.043420} {\bibfield
  {journal} {\bibinfo  {journal} {Phys. Rev. A}\ }\textbf {\bibinfo {volume}
  {88}},\ \bibinfo {pages} {043420} (\bibinfo {year} {2013})}\BibitemShut
  {NoStop}%
\bibitem [{\citenamefont {Stickler}\ \emph {et~al.}(2018)\citenamefont
  {Stickler}, \citenamefont {Schrinski},\ and\ \citenamefont
  {Hornberger}}]{stickler_rotational_2018}%
  \BibitemOpen
  \bibfield  {author} {\bibinfo {author} {\bibfnamefont {B.~A.}\ \bibnamefont
  {Stickler}}, \bibinfo {author} {\bibfnamefont {B.}~\bibnamefont {Schrinski}},
  \ and\ \bibinfo {author} {\bibfnamefont {K.}~\bibnamefont {Hornberger}},\
  }\bibfield  {title} {\enquote {\bibinfo {title} {Rotational {Friction} and
  {Diffusion} of {Quantum} {Rotors}},}\ }\href {\doibase
  10.1103/PhysRevLett.121.040401} {\bibfield  {journal} {\bibinfo  {journal}
  {Phys. Rev. Lett.}\ }\textbf {\bibinfo {volume} {121}},\ \bibinfo {pages}
  {040401} (\bibinfo {year} {2018})}\BibitemShut {NoStop}%
\bibitem [{\citenamefont {Schmidt}\ and\ \citenamefont
  {Lemeshko}(2015)}]{schmidt_rotation_2015}%
  \BibitemOpen
  \bibfield  {author} {\bibinfo {author} {\bibfnamefont {R.}~\bibnamefont
  {Schmidt}}\ and\ \bibinfo {author} {\bibfnamefont {M.}~\bibnamefont
  {Lemeshko}},\ }\bibfield  {title} {\enquote {\bibinfo {title} {Rotation of
  {Quantum} {Impurities} in the {Presence} of a {Many}-{Body} {Environment}},}\
  }\href {\doibase 10.1103/PhysRevLett.114.203001} {\bibfield  {journal}
  {\bibinfo  {journal} {Phys. Rev. Lett.}\ }\textbf {\bibinfo {volume} {114}},\
  \bibinfo {pages} {203001} (\bibinfo {year} {2015})}\BibitemShut {NoStop}%
\bibitem [{\citenamefont {Lemeshko}(2017)}]{lemeshko_quasiparticle_2017}%
  \BibitemOpen
  \bibfield  {author} {\bibinfo {author} {\bibfnamefont {M.}~\bibnamefont
  {Lemeshko}},\ }\bibfield  {title} {\enquote {\bibinfo {title} {Quasiparticle
  {Approach} to {Molecules} {Interacting} with {Quantum} {Solvents}},}\ }\href
  {\doibase 10.1103/PhysRevLett.118.095301} {\bibfield  {journal} {\bibinfo
  {journal} {Phys. Rev. Lett.}\ }\textbf {\bibinfo {volume} {118}},\ \bibinfo
  {pages} {095301} (\bibinfo {year} {2017})}\BibitemShut {NoStop}%
\bibitem [{\citenamefont {Toennies}\ and\ \citenamefont
  {Vilesov}(2004)}]{toennies_superfluid_2004}%
  \BibitemOpen
  \bibfield  {author} {\bibinfo {author} {\bibfnamefont {J.~P.}\ \bibnamefont
  {Toennies}}\ and\ \bibinfo {author} {\bibfnamefont {A.~F.}\ \bibnamefont
  {Vilesov}},\ }\bibfield  {title} {\enquote {\bibinfo {title} {Superfluid
  {Helium} {Droplets}: {A} {Uniquely} {Cold} {Nanomatrix} for {Molecules} and
  {Molecular} {Complexes}},}\ }\href {\doibase 10.1002/anie.200300611}
  {\bibfield  {journal} {\bibinfo  {journal} {Angew. Chem. Int. Ed.}\ }\textbf
  {\bibinfo {volume} {43}},\ \bibinfo {pages} {2622} (\bibinfo {year}
  {2004})}\BibitemShut {NoStop}%
\bibitem [{\citenamefont {Chatterley}\ \emph {et~al.}(2018)\citenamefont
  {Chatterley}, \citenamefont {Schouder}, \citenamefont {Christiansen},
  \citenamefont {Shepperson}, \citenamefont {Rasmussen},\ and\ \citenamefont
  {Stapelfeldt}}]{chatterley_long_2018}%
  \BibitemOpen
  \bibfield  {author} {\bibinfo {author} {\bibfnamefont {A.~S.}\ \bibnamefont
  {Chatterley}}, \bibinfo {author} {\bibfnamefont {C.}~\bibnamefont
  {Schouder}}, \bibinfo {author} {\bibfnamefont {L.}~\bibnamefont
  {Christiansen}}, \bibinfo {author} {\bibfnamefont {B.}~\bibnamefont
  {Shepperson}}, \bibinfo {author} {\bibfnamefont {M.~H.}\ \bibnamefont
  {Rasmussen}}, \ and\ \bibinfo {author} {\bibfnamefont {H.}~\bibnamefont
  {Stapelfeldt}},\ }\bibfield  {title} {\enquote {\bibinfo {title}
  {Long-lasting field-free alignment of large molecules inside helium
  nanodroplets},}\ }\href@noop {} {\bibfield  {journal} {\bibinfo  {journal}
  {ArXiv e-prints}\ } (\bibinfo {year} {2018})},\ \Eprint
  {http://arxiv.org/abs/1807.07376} {arXiv:1807.07376} \BibitemShut {NoStop}%
\bibitem [{\citenamefont {Meijer}\ \emph {et~al.}(1990)\citenamefont {Meijer},
  \citenamefont {Vries}, \citenamefont {Hunziker},\ and\ \citenamefont
  {Wendt}}]{meijer_laser_1990}%
  \BibitemOpen
  \bibfield  {author} {\bibinfo {author} {\bibfnamefont {G.}~\bibnamefont
  {Meijer}}, \bibinfo {author} {\bibfnamefont {M.~S.}\ \bibnamefont {Vries}},
  \bibinfo {author} {\bibfnamefont {H.~E.}\ \bibnamefont {Hunziker}}, \ and\
  \bibinfo {author} {\bibfnamefont {H.~R.}\ \bibnamefont {Wendt}},\ }\bibfield
  {title} {\enquote {\bibinfo {title} {Laser desorption jet-cooling of organic
  molecules},}\ }\href {\doibase 10.1007/BF00329101} {\bibfield  {journal}
  {\bibinfo  {journal} {Appl. Phys. B}\ }\textbf {\bibinfo {volume} {51}},\
  \bibinfo {pages} {395--403} (\bibinfo {year} {1990})}\BibitemShut {NoStop}%
\bibitem [{\citenamefont {Choi}\ \emph {et~al.}(2006)\citenamefont {Choi},
  \citenamefont {Douberly}, \citenamefont {Falconer}, \citenamefont {Lewis},
  \citenamefont {Lindsay}, \citenamefont {Merritt}, \citenamefont {Stiles},\
  and\ \citenamefont {Miller}}]{choi_infrared_2006}%
  \BibitemOpen
  \bibfield  {author} {\bibinfo {author} {\bibfnamefont {M.~Y.}\ \bibnamefont
  {Choi}}, \bibinfo {author} {\bibfnamefont {G.~E.}\ \bibnamefont {Douberly}},
  \bibinfo {author} {\bibfnamefont {T.~M.}\ \bibnamefont {Falconer}}, \bibinfo
  {author} {\bibfnamefont {W.~K.}\ \bibnamefont {Lewis}}, \bibinfo {author}
  {\bibfnamefont {C.~M.}\ \bibnamefont {Lindsay}}, \bibinfo {author}
  {\bibfnamefont {J.~M.}\ \bibnamefont {Merritt}}, \bibinfo {author}
  {\bibfnamefont {P.~L.}\ \bibnamefont {Stiles}}, \ and\ \bibinfo {author}
  {\bibfnamefont {R.~E.}\ \bibnamefont {Miller}},\ }\bibfield  {title}
  {\enquote {\bibinfo {title} {Infrared spectroscopy of helium nanodroplets:
  novel methods for physics and chemistry},}\ }\href {\doibase
  10.1080/01442350600625092} {\bibfield  {journal} {\bibinfo  {journal} {Int.
  Rev. Phys. Chem.}\ }\textbf {\bibinfo {volume} {25}},\ \bibinfo {pages} {15}
  (\bibinfo {year} {2006})}\BibitemShut {NoStop}%
\bibitem [{\citenamefont {Yang}\ and\ \citenamefont
  {Ellis}(2012)}]{yang_helium_2012}%
  \BibitemOpen
  \bibfield  {author} {\bibinfo {author} {\bibfnamefont {S.}~\bibnamefont
  {Yang}}\ and\ \bibinfo {author} {\bibfnamefont {A.~M.}\ \bibnamefont
  {Ellis}},\ }\bibfield  {title} {\enquote {\bibinfo {title} {Helium droplets:
  a chemistry perspective},}\ }\href {\doibase 10.1039/C2CS35277J} {\bibfield
  {journal} {\bibinfo  {journal} {Chem. Soc. Rev.}\ }\textbf {\bibinfo {volume}
  {42}},\ \bibinfo {pages} {472} (\bibinfo {year} {2012})}\BibitemShut
  {NoStop}%
\bibitem [{\citenamefont {Pickering}\ \emph
  {et~al.}(2018{\natexlab{a}})\citenamefont {Pickering}, \citenamefont
  {Shepperson}, \citenamefont {Christiansen},\ and\ \citenamefont
  {Stapelfeldt}}]{pickering_femtosecond_2018}%
  \BibitemOpen
  \bibfield  {author} {\bibinfo {author} {\bibfnamefont {J.~D.}\ \bibnamefont
  {Pickering}}, \bibinfo {author} {\bibfnamefont {B.}~\bibnamefont
  {Shepperson}}, \bibinfo {author} {\bibfnamefont {L.}~\bibnamefont
  {Christiansen}}, \ and\ \bibinfo {author} {\bibfnamefont {H.}~\bibnamefont
  {Stapelfeldt}},\ }\bibfield  {title} {\enquote {\bibinfo {title} {Femtosecond
  laser-induced coulomb explosion imaging of aligned {OCS} oligomers inside
  helium nanodroplets},}\ }\href {\doibase 10.1063/1.5049555} {\bibfield
  {journal} {\bibinfo  {journal} {J. Chem. Phys.}\ }\textbf {\bibinfo {volume}
  {149}},\ \bibinfo {pages} {154306} (\bibinfo {year}
  {2018}{\natexlab{a}})}\BibitemShut {NoStop}%
\bibitem [{\citenamefont {Pickering}\ \emph
  {et~al.}(2018{\natexlab{b}})\citenamefont {Pickering}, \citenamefont
  {Shepperson}, \citenamefont {H{\"u}bschmann}, \citenamefont {Thorning},\ and\
  \citenamefont {Stapelfeldt}}]{pickering_alignment_2018}%
  \BibitemOpen
  \bibfield  {author} {\bibinfo {author} {\bibfnamefont {J.~D.}\ \bibnamefont
  {Pickering}}, \bibinfo {author} {\bibfnamefont {B.}~\bibnamefont
  {Shepperson}}, \bibinfo {author} {\bibfnamefont {B.~A.~K.}\ \bibnamefont
  {H{\"u}bschmann}}, \bibinfo {author} {\bibfnamefont {F.}~\bibnamefont
  {Thorning}}, \ and\ \bibinfo {author} {\bibfnamefont {H.}~\bibnamefont
  {Stapelfeldt}},\ }\bibfield  {title} {\enquote {\bibinfo {title} {Alignment
  and imaging of the {$CS_2$} dimer inside helium nanodroplets},}\ }\href
  {\doibase 10.1103/PhysRevLett.120.113202} {\bibfield  {journal} {\bibinfo
  {journal} {Phys. Rev. Lett.}\ }\textbf {\bibinfo {volume} {120}},\ \bibinfo
  {pages} {1121321} (\bibinfo {year} {2018}{\natexlab{b}})}\BibitemShut
  {NoStop}%
\bibitem [{\citenamefont {Stapelfeldt}\ and\ \citenamefont
  {Seideman}(2003)}]{stapelfeldt_colloquium:_2003}%
  \BibitemOpen
  \bibfield  {author} {\bibinfo {author} {\bibfnamefont {H.}~\bibnamefont
  {Stapelfeldt}}\ and\ \bibinfo {author} {\bibfnamefont {T.}~\bibnamefont
  {Seideman}},\ }\bibfield  {title} {\enquote {\bibinfo {title} {Colloquium:
  {Aligning} molecules with strong laser pulses},}\ }\href {\doibase
  10.1103/RevModPhys.75.543} {\bibfield  {journal} {\bibinfo  {journal} {Rev.
  Mod. Phys.}\ }\textbf {\bibinfo {volume} {75}},\ \bibinfo {pages} {543}
  (\bibinfo {year} {2003})}\BibitemShut {NoStop}%
\bibitem [{\citenamefont {Burt}\ \emph {et~al.}(2018)\citenamefont {Burt},
  \citenamefont {Amini}, \citenamefont {Lee}, \citenamefont {Christiansen},
  \citenamefont {Johansen}, \citenamefont {Kobayashi}, \citenamefont
  {Pickering}, \citenamefont {Vallance}, \citenamefont {Brouard},\ and\
  \citenamefont {Stapelfeldt}}]{burt_communication_2018}%
  \BibitemOpen
  \bibfield  {author} {\bibinfo {author} {\bibfnamefont {M.}~\bibnamefont
  {Burt}}, \bibinfo {author} {\bibfnamefont {K.}~\bibnamefont {Amini}},
  \bibinfo {author} {\bibfnamefont {J.~W.~L.}\ \bibnamefont {Lee}}, \bibinfo
  {author} {\bibfnamefont {L.}~\bibnamefont {Christiansen}}, \bibinfo {author}
  {\bibfnamefont {R.~R.}\ \bibnamefont {Johansen}}, \bibinfo {author}
  {\bibfnamefont {Y.}~\bibnamefont {Kobayashi}}, \bibinfo {author}
  {\bibfnamefont {J.~D.}\ \bibnamefont {Pickering}}, \bibinfo {author}
  {\bibfnamefont {C.}~\bibnamefont {Vallance}}, \bibinfo {author}
  {\bibfnamefont {M.}~\bibnamefont {Brouard}}, \ and\ \bibinfo {author}
  {\bibfnamefont {H.}~\bibnamefont {Stapelfeldt}},\ }\bibfield  {title}
  {\enquote {\bibinfo {title} {Communication: Gas-phase structural isomer
  identification by coulomb explosion of aligned molecules},}\ }\href {\doibase
  10.1063/1.5023441} {\bibfield  {journal} {\bibinfo  {journal} {J. Chem.
  Phys}\ }\textbf {\bibinfo {volume} {148}},\ \bibinfo {pages} {091102}
  (\bibinfo {year} {2018})}\BibitemShut {NoStop}%
\bibitem [{\citenamefont {Smeenk}\ and\ \citenamefont
  {Corkum}(2013)}]{smeenk_molecular_2013}%
  \BibitemOpen
  \bibfield  {author} {\bibinfo {author} {\bibfnamefont {C.~T.~L.}\
  \bibnamefont {Smeenk}}\ and\ \bibinfo {author} {\bibfnamefont {P.~B.}\
  \bibnamefont {Corkum}},\ }\bibfield  {title} {\enquote {\bibinfo {title}
  {Molecular alignment using circularly polarized laser pulses},}\ }\href
  {\doibase 10.1088/0953-4075/46/20/201001} {\bibfield  {journal} {\bibinfo
  {journal} {J. Phys. B: At. Mol. Opt. Phys.}\ }\textbf {\bibinfo {volume}
  {46}},\ \bibinfo {pages} {201001} (\bibinfo {year} {2013})}\BibitemShut
  {NoStop}%
\bibitem [{\citenamefont {Moazzen-Ahmadi}\ and\ \citenamefont
  {McKellar}(2013)}]{moazzen-ahmadi_spectroscopy_2013}%
  \BibitemOpen
  \bibfield  {author} {\bibinfo {author} {\bibfnamefont {N.}~\bibnamefont
  {Moazzen-Ahmadi}}\ and\ \bibinfo {author} {\bibfnamefont {A.~R.~W.}\
  \bibnamefont {McKellar}},\ }\bibfield  {title} {\enquote {\bibinfo {title}
  {Spectroscopy of dimers, trimers and larger clusters of linear molecules},}\
  }\href {\doibase 10.1080/0144235X.2013.813799} {\bibfield  {journal}
  {\bibinfo  {journal} {Int. Rev. Phys. Chem.}\ }\textbf {\bibinfo {volume}
  {32}},\ \bibinfo {pages} {611} (\bibinfo {year} {2013})}\BibitemShut
  {NoStop}%
\bibitem [{\citenamefont {Rezaei}\ \emph {et~al.}(2011)\citenamefont {Rezaei},
  \citenamefont {Norooz~Oliaee}, \citenamefont {Moazzen-Ahmadi},\ and\
  \citenamefont {McKellar}}]{rezaei_spectroscopic_2011}%
  \BibitemOpen
  \bibfield  {author} {\bibinfo {author} {\bibfnamefont {M.}~\bibnamefont
  {Rezaei}}, \bibinfo {author} {\bibfnamefont {J.}~\bibnamefont
  {Norooz~Oliaee}}, \bibinfo {author} {\bibfnamefont {N.}~\bibnamefont
  {Moazzen-Ahmadi}}, \ and\ \bibinfo {author} {\bibfnamefont {A.~R.~W.}\
  \bibnamefont {McKellar}},\ }\bibfield  {title} {\enquote {\bibinfo {title}
  {Spectroscopic observation and structure of {CS}$_2$ dimer},}\ }\href
  {\doibase 10.1063/1.3578177} {\bibfield  {journal} {\bibinfo  {journal} {J.
  Chem. Phys.}\ }\textbf {\bibinfo {volume} {134}},\ \bibinfo {pages} {144306}
  (\bibinfo {year} {2011})}\BibitemShut {NoStop}%
\bibitem [{\citenamefont {Smeenk}\ \emph {et~al.}(2014)\citenamefont {Smeenk},
  \citenamefont {Arissian}, \citenamefont {Sokolov}, \citenamefont {Spanner},
  \citenamefont {Lee}, \citenamefont {Staudte}, \citenamefont {Villeneuve},\
  and\ \citenamefont {Corkum}}]{smeenk_alignment_2014}%
  \BibitemOpen
  \bibfield  {author} {\bibinfo {author} {\bibfnamefont {C. T. L.}\
  \bibnamefont {Smeenk}}, \bibinfo {author} {\bibfnamefont {L.}~\bibnamefont
  {Arissian}}, \bibinfo {author} {\bibfnamefont {A. V.}\ \bibnamefont
  {Sokolov}}, \bibinfo {author} {\bibfnamefont {M.}~\bibnamefont {Spanner}},
  \bibinfo {author} {\bibfnamefont {K. F.}\ \bibnamefont {Lee}}, \bibinfo
  {author} {\bibfnamefont {A.}~\bibnamefont {Staudte}}, \bibinfo {author}
  {\bibfnamefont {D. M.}\ \bibnamefont {Villeneuve}}, \ and\ \bibinfo
  {author} {\bibfnamefont {P.~B.}\ \bibnamefont {Corkum}},\ }\bibfield  {title}
  {\enquote {\bibinfo {title} {Alignment {Dependent} {Enhancement} of the
  {Photoelectron} {Cutoff} for {Multiphoton} {Ionization} of {Molecules}},}\
  }\href {\doibase 10.1103/PhysRevLett.112.253001} {\bibfield  {journal}
  {\bibinfo  {journal} {Phys. Rev. Lett.}\ }\textbf {\bibinfo {volume} {112}},\
  \bibinfo {pages} {253001} (\bibinfo {year} {2014})}\BibitemShut {NoStop}%
\bibitem [{\citenamefont {Seideman}\ and\ \citenamefont
  {Hamilton}(2005)}]{seideman_nonadiabatic_2005}%
  \BibitemOpen
  \bibfield  {author} {\bibinfo {author} {\bibfnamefont {T.}~\bibnamefont
  {Seideman}}\ and\ \bibinfo {author} {\bibfnamefont {E.}~\bibnamefont
  {Hamilton}},\ }\bibfield  {title} {\enquote {\bibinfo {title} {Nonadiabatic
  {Alignment} by {Intense} {Pulses}. {Concepts}, {Theory}, and {Directions}},}\
  }\href {\doibase 10.1016/S1049-250X(05)52006-8} {\bibfield  {journal}
  {\bibinfo  {journal} {Adv. At. Mol. Opt. Phys}\ }\textbf {\bibinfo {volume}
  {52}},\ \bibinfo {pages} {289--329} (\bibinfo {year} {2005})}\BibitemShut
  {NoStop}%
\bibitem [{\citenamefont {Fleischer}\ \emph {et~al.}(2012)\citenamefont
  {Fleischer}, \citenamefont {Khodorkovsky}, \citenamefont {Gershnabel},
  \citenamefont {Prior},\ and\ \citenamefont
  {Averbukh}}]{fleischer_molecular_2012}%
  \BibitemOpen
  \bibfield  {author} {\bibinfo {author} {\bibfnamefont {S.}~\bibnamefont
  {Fleischer}}, \bibinfo {author} {\bibfnamefont {Y.}~\bibnamefont
  {Khodorkovsky}}, \bibinfo {author} {\bibfnamefont {E.}~\bibnamefont
  {Gershnabel}}, \bibinfo {author} {\bibfnamefont {Y.}~\bibnamefont {Prior}}, \
  and\ \bibinfo {author} {\bibfnamefont {I.~S.}\ \bibnamefont {Averbukh}},\
  }\bibfield  {title} {\enquote {\bibinfo {title} {Molecular alignment induced
  by ultrashort laser pulses and its impact on molecular motion},}\ }\href
  {\doibase 10.100/ijch.201100161} {\bibfield  {journal} {\bibinfo  {journal}
  {Isr. J. Chem.}\ }\textbf {\bibinfo {volume} {52}},\ \bibinfo {pages}
  {414--437} (\bibinfo {year} {2012})}\BibitemShut {NoStop}%
\bibitem [{\citenamefont {Larsen}(2000)}]{larsen_laser_2000}%
  \BibitemOpen
  \bibfield  {author} {\bibinfo {author} {\bibfnamefont {Jakob~Juul}\
  \bibnamefont {Larsen}},\ }\emph {\bibinfo {title} {Laser {Induced}
  {Alignment} of {Neutral} {Molecules}}},\ \href@noop {} {Ph.D. thesis},\
  \bibinfo  {school} {Aarhus University} (\bibinfo {year} {2000})\BibitemShut
  {NoStop}%
\bibitem [{\citenamefont {Søndergaard}\ \emph {et~al.}(2017)\citenamefont
  {Søndergaard}, \citenamefont {Shepperson},\ and\ \citenamefont
  {Stapelfeldt}}]{sondergaard_nonadiabatic_2017}%
  \BibitemOpen
  \bibfield  {author} {\bibinfo {author} {\bibfnamefont {A.~A.}\ \bibnamefont
  {Søndergaard}}, \bibinfo {author} {\bibfnamefont {B.}~\bibnamefont
  {Shepperson}}, \ and\ \bibinfo {author} {\bibfnamefont {H.}~\bibnamefont
  {Stapelfeldt}},\ }\bibfield  {title} {\enquote {\bibinfo {title}
  {Nonadiabatic laser-induced alignment of molecules: {Reconstructing}
  {$\langle \cos^2 \theta \rangle$} directly from {$\langle \cos^2 \theta_{2D}
  \rangle$} by {Fourier} analysis},}\ }\href {\doibase 10.1063/1.4975817}
  {\bibfield  {journal} {\bibinfo  {journal} {J. Chem. Phys.}\ }\textbf
  {\bibinfo {volume} {147}},\ \bibinfo {pages} {013905} (\bibinfo {year}
  {2017})}\BibitemShut {NoStop}%
\bibitem [{\citenamefont {Friedrich}\ and\ \citenamefont
  {Herschbach}(1995)}]{friedrich_alignment_1995}%
  \BibitemOpen
  \bibfield  {author} {\bibinfo {author} {\bibfnamefont {B.}~\bibnamefont
  {Friedrich}}\ and\ \bibinfo {author} {\bibfnamefont {D.}~\bibnamefont
  {Herschbach}},\ }\bibfield  {title} {\enquote {\bibinfo {title} {Alignment
  and trapping of molecules in intense laser fields},}\ }\href {\doibase
  10.1103/PhysRevLett.74.4623} {\bibfield  {journal} {\bibinfo  {journal}
  {Phys. Rev. Lett.}\ }\textbf {\bibinfo {volume} {74}},\ \bibinfo {pages}
  {4623--4626} (\bibinfo {year} {1995})}\BibitemShut {NoStop}%
\bibitem [{\citenamefont
  {Søndergaard}(2016)}]{sondergaard_understanding_2016}%
  \BibitemOpen
  \bibfield  {author} {\bibinfo {author} {\bibfnamefont {A.~A.}\ \bibnamefont
  {Søndergaard}},\ }\emph {\bibinfo {title} {Understanding {Laser}-{Induced}
  {Alignment} and {Rotation} of {Molecules} {Embedded} in {Helium}
  {Nanodroplets}}},\ \href@noop {} {\bibinfo {type} {Thesis}},\ \bibinfo
  {school} {Aarhus University} (\bibinfo {year} {2016})\BibitemShut {NoStop}%
\bibitem [{\citenamefont {Zillich}()}]{zillich_private_2018}%
  \BibitemOpen
  \bibfield  {author} {\bibinfo {author} {\bibfnamefont {R.}~\bibnamefont
  {Zillich}},\ }\href@noop {} {}\bibinfo {howpublished} {Private
  Communication}\BibitemShut {NoStop}%
\bibitem [{\citenamefont {Seideman}(2001)}]{seideman_dynamics_2001}%
  \BibitemOpen
  \bibfield  {author} {\bibinfo {author} {\bibfnamefont {T.}~\bibnamefont
  {Seideman}},\ }\bibfield  {title} {\enquote {\bibinfo {title} {On the
  dynamics of rotationally broad, spatially aligned wave packets},}\ }\href
  {\doibase 10.1063/1.1400131} {\bibfield  {journal} {\bibinfo  {journal} {J.
  Chem. Phys.}\ }\textbf {\bibinfo {volume} {115}},\ \bibinfo {pages}
  {5965--5973} (\bibinfo {year} {2001})}\BibitemShut {NoStop}%
\bibitem [{\citenamefont {Hansen}\ \emph {et~al.}(2012)\citenamefont {Hansen},
  \citenamefont {Nielsen}, \citenamefont {Madsen}, \citenamefont {Lindhardt},
  \citenamefont {Johansson}, \citenamefont {Skrydstrup}, \citenamefont
  {Madsen},\ and\ \citenamefont {Stapelfeldt}}]{hansen_control_2012}%
  \BibitemOpen
  \bibfield  {author} {\bibinfo {author} {\bibfnamefont {J.~L.}\ \bibnamefont
  {Hansen}}, \bibinfo {author} {\bibfnamefont {J.~H.}\ \bibnamefont {Nielsen}},
  \bibinfo {author} {\bibfnamefont {C.~B.}\ \bibnamefont {Madsen}}, \bibinfo
  {author} {\bibfnamefont {A.~T.}\ \bibnamefont {Lindhardt}}, \bibinfo {author}
  {\bibfnamefont {M.~P}\ \bibnamefont {Johansson}}, \bibinfo {author}
  {\bibfnamefont {T.}~\bibnamefont {Skrydstrup}}, \bibinfo {author}
  {\bibfnamefont {L.~B.}\ \bibnamefont {Madsen}}, \ and\ \bibinfo {author}
  {\bibfnamefont {H.}~\bibnamefont {Stapelfeldt}},\ }\bibfield  {title}
  {\enquote {\bibinfo {title} {Control and femtosecond time-resolved imaging of
  torsion in a chiral molecule},}\ }\href {\doibase 10.1063/1.4719816}
  {\bibfield  {journal} {\bibinfo  {journal} {J. Chem. Phys.}\ }\textbf
  {\bibinfo {volume} {136}},\ \bibinfo {pages} {204310} (\bibinfo {year}
  {2012})}\BibitemShut {NoStop}%
\bibitem [{\citenamefont {Christensen}\ \emph {et~al.}(2016)\citenamefont
  {Christensen}, \citenamefont {Christiansen}, \citenamefont {Shepperson},\
  and\ \citenamefont {Stapelfeldt}}]{christensen_deconvoluting_2016}%
  \BibitemOpen
  \bibfield  {author} {\bibinfo {author} {\bibfnamefont {L.}~\bibnamefont
  {Christensen}}, \bibinfo {author} {\bibfnamefont {L.}~\bibnamefont
  {Christiansen}}, \bibinfo {author} {\bibfnamefont {B.}~\bibnamefont
  {Shepperson}}, \ and\ \bibinfo {author} {\bibfnamefont {H.}~\bibnamefont
  {Stapelfeldt}},\ }\bibfield  {title} {\enquote {\bibinfo {title}
  {Deconvoluting nonaxial recoil in {Coulomb} explosion measurements of
  molecular axis alignment},}\ }\href {\doibase 10.1103/PhysRevA.94.023410}
  {\bibfield  {journal} {\bibinfo  {journal} {Phys. Rev. A}\ }\textbf {\bibinfo
  {volume} {94}},\ \bibinfo {pages} {023410} (\bibinfo {year}
  {2016})}\BibitemShut {NoStop}%
\bibitem [{\citenamefont {Hobza}\ and\ \citenamefont
  {Dethlefs-M{\"u}ller}(2010)}]{hobza_noncovalent_2010}%
  \BibitemOpen
  \bibfield  {author} {\bibinfo {author} {\bibfnamefont {P.}~\bibnamefont
  {Hobza}}\ and\ \bibinfo {author} {\bibfnamefont {K.}~\bibnamefont
  {Dethlefs-M{\"u}ller}},\ }\href@noop {} {\emph {\bibinfo {title}
  {Non-covalent Interactions, Theory and Experiment}}}\ (\bibinfo  {publisher}
  {RSC Publishing},\ \bibinfo {year} {2010})\BibitemShut {NoStop}%
\bibitem [{\citenamefont {Küpper}\ \emph {et~al.}(2014)\citenamefont
  {Küpper}, \citenamefont {Stern}, \citenamefont {Holmegaard}, \citenamefont
  {Filsinger}, \citenamefont {Rouzée}, \citenamefont {Rudenko}, \citenamefont
  {Johnsson}, \citenamefont {Martin}, \citenamefont {Adolph}, \citenamefont
  {Aquila}, \citenamefont {Bajt}, \citenamefont {Barty}, \citenamefont
  {Bostedt}, \citenamefont {Bozek}, \citenamefont {Caleman}, \citenamefont
  {Coffee}, \citenamefont {Coppola}, \citenamefont {Delmas}, \citenamefont
  {Epp}, \citenamefont {Erk}, \citenamefont {Foucar}, \citenamefont
  {Gorkhover}, \citenamefont {Gumprecht}, \citenamefont {Hartmann},
  \citenamefont {Hartmann}, \citenamefont {Hauser}, \citenamefont {Holl},
  \citenamefont {Hömke}, \citenamefont {Kimmel}, \citenamefont {Krasniqi},
  \citenamefont {Kühnel}, \citenamefont {Maurer}, \citenamefont
  {Messerschmidt}, \citenamefont {Moshammer}, \citenamefont {Reich},
  \citenamefont {Rudek}, \citenamefont {Santra}, \citenamefont {Schlichting},
  \citenamefont {Schmidt}, \citenamefont {Schorb}, \citenamefont {Schulz},
  \citenamefont {Soltau}, \citenamefont {Spence}, \citenamefont {Starodub},
  \citenamefont {Strüder}, \citenamefont {Thøgersen}, \citenamefont
  {Vrakking}, \citenamefont {Weidenspointner}, \citenamefont {White},
  \citenamefont {Wunderer}, \citenamefont {Meijer}, \citenamefont {Ullrich},
  \citenamefont {Stapelfeldt}, \citenamefont {Rolles},\ and\ \citenamefont
  {Chapman}}]{kupper_x-ray_2014}%
  \BibitemOpen
  \bibfield  {author} {\bibinfo {author} {\bibfnamefont {Jochen}\ \bibnamefont
  {Küpper}}, \bibinfo {author} {\bibfnamefont {Stephan}\ \bibnamefont
  {Stern}}, \bibinfo {author} {\bibfnamefont {Lotte}\ \bibnamefont
  {Holmegaard}}, \bibinfo {author} {\bibfnamefont {Frank}\ \bibnamefont
  {Filsinger}}, \bibinfo {author} {\bibfnamefont {Arnaud}\ \bibnamefont
  {Rouzée}}, \bibinfo {author} {\bibfnamefont {Artem}\ \bibnamefont
  {Rudenko}}, \bibinfo {author} {\bibfnamefont {Per}\ \bibnamefont {Johnsson}},
  \bibinfo {author} {\bibfnamefont {Andrew~V.}\ \bibnamefont {Martin}},
  \bibinfo {author} {\bibfnamefont {Marcus}\ \bibnamefont {Adolph}}, \bibinfo
  {author} {\bibfnamefont {Andrew}\ \bibnamefont {Aquila}}, \bibinfo {author}
  {\bibfnamefont {Saša}\ \bibnamefont {Bajt}}, \bibinfo {author}
  {\bibfnamefont {Anton}\ \bibnamefont {Barty}}, \bibinfo {author}
  {\bibfnamefont {Christoph}\ \bibnamefont {Bostedt}}, \bibinfo {author}
  {\bibfnamefont {John}\ \bibnamefont {Bozek}}, \bibinfo {author}
  {\bibfnamefont {Carl}\ \bibnamefont {Caleman}}, \bibinfo {author}
  {\bibfnamefont {Ryan}\ \bibnamefont {Coffee}}, \bibinfo {author}
  {\bibfnamefont {Nicola}\ \bibnamefont {Coppola}}, \bibinfo {author}
  {\bibfnamefont {Tjark}\ \bibnamefont {Delmas}}, \bibinfo {author}
  {\bibfnamefont {Sascha}\ \bibnamefont {Epp}}, \bibinfo {author}
  {\bibfnamefont {Benjamin}\ \bibnamefont {Erk}}, \bibinfo {author}
  {\bibfnamefont {Lutz}\ \bibnamefont {Foucar}}, \bibinfo {author}
  {\bibfnamefont {Tais}\ \bibnamefont {Gorkhover}}, \bibinfo {author}
  {\bibfnamefont {Lars}\ \bibnamefont {Gumprecht}}, \bibinfo {author}
  {\bibfnamefont {Andreas}\ \bibnamefont {Hartmann}}, \bibinfo {author}
  {\bibfnamefont {Robert}\ \bibnamefont {Hartmann}}, \bibinfo {author}
  {\bibfnamefont {Günter}\ \bibnamefont {Hauser}}, \bibinfo {author}
  {\bibfnamefont {Peter}\ \bibnamefont {Holl}}, \bibinfo {author}
  {\bibfnamefont {Andre}\ \bibnamefont {Hömke}}, \bibinfo {author}
  {\bibfnamefont {Nils}\ \bibnamefont {Kimmel}}, \bibinfo {author}
  {\bibfnamefont {Faton}\ \bibnamefont {Krasniqi}}, \bibinfo {author}
  {\bibfnamefont {Kai-Uwe}\ \bibnamefont {Kühnel}}, \bibinfo {author}
  {\bibfnamefont {Jochen}\ \bibnamefont {Maurer}}, \bibinfo {author}
  {\bibfnamefont {Marc}\ \bibnamefont {Messerschmidt}}, \bibinfo {author}
  {\bibfnamefont {Robert}\ \bibnamefont {Moshammer}}, \bibinfo {author}
  {\bibfnamefont {Christian}\ \bibnamefont {Reich}}, \bibinfo {author}
  {\bibfnamefont {Benedikt}\ \bibnamefont {Rudek}}, \bibinfo {author}
  {\bibfnamefont {Robin}\ \bibnamefont {Santra}}, \bibinfo {author}
  {\bibfnamefont {Ilme}\ \bibnamefont {Schlichting}}, \bibinfo {author}
  {\bibfnamefont {Carlo}\ \bibnamefont {Schmidt}}, \bibinfo {author}
  {\bibfnamefont {Sebastian}\ \bibnamefont {Schorb}}, \bibinfo {author}
  {\bibfnamefont {Joachim}\ \bibnamefont {Schulz}}, \bibinfo {author}
  {\bibfnamefont {Heike}\ \bibnamefont {Soltau}}, \bibinfo {author}
  {\bibfnamefont {John C.~H.}\ \bibnamefont {Spence}}, \bibinfo {author}
  {\bibfnamefont {Dmitri}\ \bibnamefont {Starodub}}, \bibinfo {author}
  {\bibfnamefont {Lothar}\ \bibnamefont {Strüder}}, \bibinfo {author}
  {\bibfnamefont {Jan}\ \bibnamefont {Thøgersen}}, \bibinfo {author}
  {\bibfnamefont {Marc J.~J.}\ \bibnamefont {Vrakking}}, \bibinfo {author}
  {\bibfnamefont {Georg}\ \bibnamefont {Weidenspointner}}, \bibinfo {author}
  {\bibfnamefont {Thomas~A.}\ \bibnamefont {White}}, \bibinfo {author}
  {\bibfnamefont {Cornelia}\ \bibnamefont {Wunderer}}, \bibinfo {author}
  {\bibfnamefont {Gerard}\ \bibnamefont {Meijer}}, \bibinfo {author}
  {\bibfnamefont {Joachim}\ \bibnamefont {Ullrich}}, \bibinfo {author}
  {\bibfnamefont {Henrik}\ \bibnamefont {Stapelfeldt}}, \bibinfo {author}
  {\bibfnamefont {Daniel}\ \bibnamefont {Rolles}}, \ and\ \bibinfo {author}
  {\bibfnamefont {Henry~N.}\ \bibnamefont {Chapman}},\ }\bibfield  {title}
  {\enquote {\bibinfo {title} {X-{Ray} {Diffraction} from {Isolated} and
  {Strongly} {Aligned} {Gas}-{Phase} {Molecules} with a {Free}-{Electron}
  {Laser}},}\ }\href {\doibase 10.1103/PhysRevLett.112.083002} {\bibfield
  {journal} {\bibinfo  {journal} {Phys. Rev. Lett.}\ }\textbf {\bibinfo
  {volume} {112}},\ \bibinfo {pages} {083002} (\bibinfo {year}
  {2014})}\BibitemShut {NoStop}%
\bibitem [{\citenamefont {Gomez}\ \emph {et~al.}(2014)\citenamefont {Gomez},
  \citenamefont {Ferguson}, \citenamefont {Cryan}, \citenamefont {Bacellar},
  \citenamefont {Tanyag}, \citenamefont {Jones}, \citenamefont {Schorb},
  \citenamefont {Anielski}, \citenamefont {Belkacem}, \citenamefont {Bernando},
  \citenamefont {Boll}, \citenamefont {Bozek}, \citenamefont {Carron},
  \citenamefont {Chen}, \citenamefont {Delmas}, \citenamefont {Englert},
  \citenamefont {Epp}, \citenamefont {Erk}, \citenamefont {Foucar},
  \citenamefont {Hartmann}, \citenamefont {Hexemer}, \citenamefont {Huth},
  \citenamefont {Kwok}, \citenamefont {Leone}, \citenamefont {Ma},
  \citenamefont {Maia}, \citenamefont {Malmerberg}, \citenamefont {Marchesini},
  \citenamefont {Neumark}, \citenamefont {Poon}, \citenamefont {Prell},
  \citenamefont {Rolles}, \citenamefont {Rudek}, \citenamefont {Rudenko},
  \citenamefont {Seifrid}, \citenamefont {Siefermann}, \citenamefont {Sturm},
  \citenamefont {Swiggers}, \citenamefont {Ullrich}, \citenamefont {Weise},
  \citenamefont {Zwart}, \citenamefont {Bostedt}, \citenamefont {Gessner},\
  and\ \citenamefont {Vilesov}}]{gomez_shapes_2014}%
  \BibitemOpen
  \bibfield  {author} {\bibinfo {author} {\bibfnamefont {Luis~F.}\ \bibnamefont
  {Gomez}}, \bibinfo {author} {\bibfnamefont {Ken~R.}\ \bibnamefont
  {Ferguson}}, \bibinfo {author} {\bibfnamefont {James~P.}\ \bibnamefont
  {Cryan}}, \bibinfo {author} {\bibfnamefont {Camila}\ \bibnamefont
  {Bacellar}}, \bibinfo {author} {\bibfnamefont {Rico Mayro~P.}\ \bibnamefont
  {Tanyag}}, \bibinfo {author} {\bibfnamefont {Curtis}\ \bibnamefont {Jones}},
  \bibinfo {author} {\bibfnamefont {Sebastian}\ \bibnamefont {Schorb}},
  \bibinfo {author} {\bibfnamefont {Denis}\ \bibnamefont {Anielski}}, \bibinfo
  {author} {\bibfnamefont {Ali}\ \bibnamefont {Belkacem}}, \bibinfo {author}
  {\bibfnamefont {Charles}\ \bibnamefont {Bernando}}, \bibinfo {author}
  {\bibfnamefont {Rebecca}\ \bibnamefont {Boll}}, \bibinfo {author}
  {\bibfnamefont {John}\ \bibnamefont {Bozek}}, \bibinfo {author}
  {\bibfnamefont {Sebastian}\ \bibnamefont {Carron}}, \bibinfo {author}
  {\bibfnamefont {Gang}\ \bibnamefont {Chen}}, \bibinfo {author} {\bibfnamefont
  {Tjark}\ \bibnamefont {Delmas}}, \bibinfo {author} {\bibfnamefont {Lars}\
  \bibnamefont {Englert}}, \bibinfo {author} {\bibfnamefont {Sascha~W.}\
  \bibnamefont {Epp}}, \bibinfo {author} {\bibfnamefont {Benjamin}\
  \bibnamefont {Erk}}, \bibinfo {author} {\bibfnamefont {Lutz}\ \bibnamefont
  {Foucar}}, \bibinfo {author} {\bibfnamefont {Robert}\ \bibnamefont
  {Hartmann}}, \bibinfo {author} {\bibfnamefont {Alexander}\ \bibnamefont
  {Hexemer}}, \bibinfo {author} {\bibfnamefont {Martin}\ \bibnamefont {Huth}},
  \bibinfo {author} {\bibfnamefont {Justin}\ \bibnamefont {Kwok}}, \bibinfo
  {author} {\bibfnamefont {Stephen~R.}\ \bibnamefont {Leone}}, \bibinfo
  {author} {\bibfnamefont {Jonathan H.~S.}\ \bibnamefont {Ma}}, \bibinfo
  {author} {\bibfnamefont {Filipe R. N.~C.}\ \bibnamefont {Maia}}, \bibinfo
  {author} {\bibfnamefont {Erik}\ \bibnamefont {Malmerberg}}, \bibinfo {author}
  {\bibfnamefont {Stefano}\ \bibnamefont {Marchesini}}, \bibinfo {author}
  {\bibfnamefont {Daniel~M.}\ \bibnamefont {Neumark}}, \bibinfo {author}
  {\bibfnamefont {Billy}\ \bibnamefont {Poon}}, \bibinfo {author}
  {\bibfnamefont {James}\ \bibnamefont {Prell}}, \bibinfo {author}
  {\bibfnamefont {Daniel}\ \bibnamefont {Rolles}}, \bibinfo {author}
  {\bibfnamefont {Benedikt}\ \bibnamefont {Rudek}}, \bibinfo {author}
  {\bibfnamefont {Artem}\ \bibnamefont {Rudenko}}, \bibinfo {author}
  {\bibfnamefont {Martin}\ \bibnamefont {Seifrid}}, \bibinfo {author}
  {\bibfnamefont {Katrin~R.}\ \bibnamefont {Siefermann}}, \bibinfo {author}
  {\bibfnamefont {Felix~P.}\ \bibnamefont {Sturm}}, \bibinfo {author}
  {\bibfnamefont {Michele}\ \bibnamefont {Swiggers}}, \bibinfo {author}
  {\bibfnamefont {Joachim}\ \bibnamefont {Ullrich}}, \bibinfo {author}
  {\bibfnamefont {Fabian}\ \bibnamefont {Weise}}, \bibinfo {author}
  {\bibfnamefont {Petrus}\ \bibnamefont {Zwart}}, \bibinfo {author}
  {\bibfnamefont {Christoph}\ \bibnamefont {Bostedt}}, \bibinfo {author}
  {\bibfnamefont {Oliver}\ \bibnamefont {Gessner}}, \ and\ \bibinfo {author}
  {\bibfnamefont {Andrey~F.}\ \bibnamefont {Vilesov}},\ }\bibfield  {title}
  {\enquote {\bibinfo {title} {Shapes and vorticities of superfluid helium
  nanodroplets},}\ }\href {\doibase 10.1126/science.1252395} {\bibfield
  {journal} {\bibinfo  {journal} {Science}\ }\textbf {\bibinfo {volume}
  {345}},\ \bibinfo {pages} {906--909} (\bibinfo {year} {2014})}\BibitemShut
  {NoStop}%
\bibitem [{\citenamefont {Tanyag}\ \emph {et~al.}(2015)\citenamefont {Tanyag},
  \citenamefont {Bernando}, \citenamefont {Jones}, \citenamefont {Bacellar},
  \citenamefont {Ferguson}, \citenamefont {Anielski}, \citenamefont {Boll},
  \citenamefont {Carron}, \citenamefont {Cryan}, \citenamefont {Englert},
  \citenamefont {Epp}, \citenamefont {Erk}, \citenamefont {Foucar},
  \citenamefont {Gomez}, \citenamefont {Hartmann}, \citenamefont {Neumark},
  \citenamefont {Rolles}, \citenamefont {Rudek}, \citenamefont {Rudenko},
  \citenamefont {Siefermann}, \citenamefont {Ullrich}, \citenamefont {Weise},
  \citenamefont {Bostedt}, \citenamefont {Gessner},\ and\ \citenamefont
  {Vilesov}}]{tanyag_communication:_2015}%
  \BibitemOpen
  \bibfield  {author} {\bibinfo {author} {\bibfnamefont {Rico Mayro~P.}\
  \bibnamefont {Tanyag}}, \bibinfo {author} {\bibfnamefont {Charles}\
  \bibnamefont {Bernando}}, \bibinfo {author} {\bibfnamefont {Curtis~F.}\
  \bibnamefont {Jones}}, \bibinfo {author} {\bibfnamefont {Camila}\
  \bibnamefont {Bacellar}}, \bibinfo {author} {\bibfnamefont {Ken~R.}\
  \bibnamefont {Ferguson}}, \bibinfo {author} {\bibfnamefont {Denis}\
  \bibnamefont {Anielski}}, \bibinfo {author} {\bibfnamefont {Rebecca}\
  \bibnamefont {Boll}}, \bibinfo {author} {\bibfnamefont {Sebastian}\
  \bibnamefont {Carron}}, \bibinfo {author} {\bibfnamefont {James~P.}\
  \bibnamefont {Cryan}}, \bibinfo {author} {\bibfnamefont {Lars}\ \bibnamefont
  {Englert}}, \bibinfo {author} {\bibfnamefont {Sascha~W.}\ \bibnamefont
  {Epp}}, \bibinfo {author} {\bibfnamefont {Benjamin}\ \bibnamefont {Erk}},
  \bibinfo {author} {\bibfnamefont {Lutz}\ \bibnamefont {Foucar}}, \bibinfo
  {author} {\bibfnamefont {Luis~F.}\ \bibnamefont {Gomez}}, \bibinfo {author}
  {\bibfnamefont {Robert}\ \bibnamefont {Hartmann}}, \bibinfo {author}
  {\bibfnamefont {Daniel~M.}\ \bibnamefont {Neumark}}, \bibinfo {author}
  {\bibfnamefont {Daniel}\ \bibnamefont {Rolles}}, \bibinfo {author}
  {\bibfnamefont {Benedikt}\ \bibnamefont {Rudek}}, \bibinfo {author}
  {\bibfnamefont {Artem}\ \bibnamefont {Rudenko}}, \bibinfo {author}
  {\bibfnamefont {Katrin~R.}\ \bibnamefont {Siefermann}}, \bibinfo {author}
  {\bibfnamefont {Joachim}\ \bibnamefont {Ullrich}}, \bibinfo {author}
  {\bibfnamefont {Fabian}\ \bibnamefont {Weise}}, \bibinfo {author}
  {\bibfnamefont {Christoph}\ \bibnamefont {Bostedt}}, \bibinfo {author}
  {\bibfnamefont {Oliver}\ \bibnamefont {Gessner}}, \ and\ \bibinfo {author}
  {\bibfnamefont {Andrey~F.}\ \bibnamefont {Vilesov}},\ }\bibfield  {title}
  {\enquote {\bibinfo {title} {Communication: {X}-ray coherent diffractive
  imaging by immersion in nanodroplets},}\ }\href {\doibase 10.1063/1.4933297}
  {\bibfield  {journal} {\bibinfo  {journal} {Struc. Dyn.}\ }\textbf {\bibinfo
  {volume} {2}},\ \bibinfo {pages} {051102} (\bibinfo {year}
  {2015})}\BibitemShut {NoStop}%
\bibitem [{\citenamefont {Rupp}\ \emph {et~al.}(2017)\citenamefont {Rupp},
  \citenamefont {Monserud}, \citenamefont {Langbehn}, \citenamefont {Sauppe},
  \citenamefont {Zimmermann}, \citenamefont {Ovcharenko}, \citenamefont
  {M{\"o}ller}, \citenamefont {Frassetto}, \citenamefont {Poletto},
  \citenamefont {Trabattoni}, \citenamefont {Calegari}, \citenamefont {Nisoli},
  \citenamefont {Sander}, \citenamefont {Peltz}, \citenamefont {Vrakking},
  \citenamefont {Fennel},\ and\ \citenamefont
  {Rouz{\'e}e}}]{rupp_coherent_2017}%
  \BibitemOpen
  \bibfield  {author} {\bibinfo {author} {\bibfnamefont {D.}~\bibnamefont
  {Rupp}}, \bibinfo {author} {\bibfnamefont {N.}~\bibnamefont {Monserud}},
  \bibinfo {author} {\bibfnamefont {B.}~\bibnamefont {Langbehn}}, \bibinfo
  {author} {\bibfnamefont {M.}~\bibnamefont {Sauppe}}, \bibinfo {author}
  {\bibfnamefont {J.}~\bibnamefont {Zimmermann}}, \bibinfo {author}
  {\bibfnamefont {Y.}~\bibnamefont {Ovcharenko}}, \bibinfo {author}
  {\bibfnamefont {T.}~\bibnamefont {M{\"o}ller}}, \bibinfo {author}
  {\bibfnamefont {F.}~\bibnamefont {Frassetto}}, \bibinfo {author}
  {\bibfnamefont {L.}~\bibnamefont {Poletto}}, \bibinfo {author} {\bibfnamefont
  {A.}~\bibnamefont {Trabattoni}}, \bibinfo {author} {\bibfnamefont
  {F.}~\bibnamefont {Calegari}}, \bibinfo {author} {\bibfnamefont
  {M.}~\bibnamefont {Nisoli}}, \bibinfo {author} {\bibfnamefont
  {K.}~\bibnamefont {Sander}}, \bibinfo {author} {\bibfnamefont
  {C.}~\bibnamefont {Peltz}}, \bibinfo {author} {\bibfnamefont {M.~J.}\
  \bibnamefont {Vrakking}}, \bibinfo {author} {\bibfnamefont {T.}~\bibnamefont
  {Fennel}}, \ and\ \bibinfo {author} {\bibfnamefont {A.}~\bibnamefont
  {Rouz{\'e}e}},\ }\bibfield  {title} {\enquote {\bibinfo {title} {Coherent
  diffractive imaging of single helium nanodroplets with a high harmonic
  generation source},}\ }\href {\doibase 10.1038/s41467-017-00287-z} {\bibfield
   {journal} {\bibinfo  {journal} {Nat. Comms.}\ }\textbf {\bibinfo {volume}
  {8}},\ \bibinfo {pages} {493} (\bibinfo {year} {2017})}\BibitemShut {NoStop}%
\bibitem [{\citenamefont {He}\ \emph {et~al.}(2017)\citenamefont {He},
  \citenamefont {Zhang}, \citenamefont {Lei},\ and\ \citenamefont
  {Kong}}]{he_self_2017}%
  \BibitemOpen
  \bibfield  {author} {\bibinfo {author} {\bibfnamefont {Y.}~\bibnamefont
  {He}}, \bibinfo {author} {\bibfnamefont {J.}~\bibnamefont {Zhang}}, \bibinfo
  {author} {\bibfnamefont {L.}~\bibnamefont {Lei}}, \ and\ \bibinfo {author}
  {\bibfnamefont {W.}~\bibnamefont {Kong}},\ }\bibfield  {title} {\enquote
  {\bibinfo {title} {Self-assembly of iodine in superfluid helium droplets:
  Halogen bonds and nanocrystals},}\ }\href {\doibase 10.1002/ange.201611922}
  {\bibfield  {journal} {\bibinfo  {journal} {Angw. Chem.}\ }\textbf {\bibinfo
  {volume} {129}},\ \bibinfo {pages} {3595--3599} (\bibinfo {year}
  {2017})}\BibitemShut {NoStop}%
\bibitem [{\citenamefont {Zhang}\ \emph {et~al.}(2014)\citenamefont {Zhang},
  \citenamefont {He}, \citenamefont {Freund},\ and\ \citenamefont
  {Kong}}]{zhang_electron_2014}%
  \BibitemOpen
  \bibfield  {author} {\bibinfo {author} {\bibfnamefont {J.}~\bibnamefont
  {Zhang}}, \bibinfo {author} {\bibfnamefont {Y.}~\bibnamefont {He}}, \bibinfo
  {author} {\bibfnamefont {W.~M.}\ \bibnamefont {Freund}}, \ and\ \bibinfo
  {author} {\bibfnamefont {W.}~\bibnamefont {Kong}},\ }\bibfield  {title}
  {\enquote {\bibinfo {title} {Electron {Diffraction} of {Superfluid} {Helium}
  {Droplets}},}\ }\href {\doibase 10.1021/jz5006829} {\bibfield  {journal}
  {\bibinfo  {journal} {J. Phys. Chem. Lett.}\ }\textbf {\bibinfo {volume}
  {5}},\ \bibinfo {pages} {1801--1805} (\bibinfo {year} {2014})}\BibitemShut
  {NoStop}%
\end{thebibliography}

%

\end{document}